\documentclass[twocolumn,nofootinbib,superscriptaddress]{revtex4-1}
\usepackage{latexsym,epsfig,amssymb, amsmath,nicefrac}

\usepackage{color}
\usepackage[makeroom]{cancel}
  \usepackage{latexsym}
  \usepackage{epsf}
  \usepackage{amssymb}
  \usepackage{graphicx}
  \usepackage{amsmath}
  \usepackage{amsmath,amssymb,amsthm}
  \usepackage{verbatim}
  \usepackage{hyperref}

\newcommand{\e}{\textrm{e}}

\def\tb0{\tilde{\beta}_0}
{\def\b0{\beta_0}

\def\bi{\begin{itemize}}
\def\ei{\end{itemize}}
\def\be{\begin{equation}}
\def\ee{\end{equation}}
\newcommand{\bea}{\begin{eqnarray}}
\newcommand{\eea}{\end{eqnarray}}
 
\renewcommand{\Im}{\textrm{Im}\,}
\renewcommand{\Re}{\textrm{Re}\,}

\def\Kahler{K\"{a}hler~}

\begin{document}

\vspace{1cm}

\title{Cosmological Attractors from  $\alpha$-Scale Supergravity}

\author{Diederik Roest}
\email{d.roest@rug.nl}
\author{Marco Scalisi}
\email{m.scalisi@rug.nl}
\affiliation{Van Swinderen Institute for Particle Physics and Gravity, University of Groningen, \\ Nijenborgh 4, 9747 AG Groningen, The Netherlands}
%

\begin{abstract}
The Planck value of the spectral index can be interpreted as $n_s = 1 - 2/N$ in terms of the number of e-foldings $N$. An appealing explanation for this phenomenological observation is provided by $\alpha$-attractors: the inflationary predictions of these supergravity models are fully determined by the curvature of the \Kahler manifold. We provide a novel formulation of $\alpha$-attractors which only involves a single chiral superfield. Our construction involves a natural deformation of no-scale models, and employs these to construct a De Sitter plateau with an exponential fall-off. Finally, we show how analogous structures with a flat \Kahler geometry arise as a singular limit of such $\alpha$-scale models.

\end{abstract}	

\maketitle

\section{Introduction}

The Planck satellite has found unequivocal evidence for a key prediction of inflation: the spectrum of primordial scalar perturbations, whose imprint as temperature fluctuations on the CMB was first measured by COBE \cite{COBE}, has a small deviation from scale invariance. This is encoded in the spectral index, whose value has been measured to be $n_s = 0.968\pm 0.006$ \cite{Planck:2015xua,Ade:2015lrj}, where scale invariance corresponds to $n_s = 1$. 

The particular value of the tilt has invited suggestions that this can be modelled as 
 \begin{align} \label{ns}
   n_s = 1- \frac2N \,,
  \end{align}  
possibly up to higher-order corrections in $1/N$, where $N$ is the number of e-folds between horizon crossing of the CMB modes and the end of inflation. Interestingly, this relation generically allows for only two regimes of the tensor-to-scalar ratio $r$, which is either given by $r =8/N$ or scales as $1/N^2$ at lowest order \cite{Mukhanov:2013tua, Roest:2013fha} (see also \cite{Garcia-Bellido:2014gna,Creminelli:2014nqa,Zavala:2014bda}). Furthermore, assuming eq.~\eqref{ns} strengthens the usual Lyth bound \cite{Lyth:1996im,Boubekeur:2005zm} on the displacement of the inflaton field by two orders of magnitude \cite{Garcia-Bellido:2014wfa} .

In addition to observations, there is also a theoretical predisposition for the specific scaling \eqref{ns} rather than with other coefficients of the $1/N$ term.  First examples include the Starobinsky model \cite{Starobinsky:1980te}, its supergravity implementations \cite{Cecotti:1987sa, Ellis:2013xoa,Kallosh:2013lkr,Buchmuller:2013zfa,Farakos:2013cqa,Ellis:2013nxa} and Higgs inflation \cite{Bezrukov:2007ep}, which have $r = 12 /N^2$. Moreover, it was later realized that a large set of models with arbitrary scalar potentials and specific non-minimal coupling lead to the same predictions, a phenomenon referred to as the universal attractor \cite{Kallosh:2013tua}. 

Prior to this, it had been pointed out that the same insensitivity to the details of the scalar potential arises in supergravity models with a non-trivial \Kahler geometry \cite{Kallosh:2013hoa}. The latter induces a boundary in moduli space, where the theory attains a conformal symmetry. Inflation takes places as the inflaton moves away from the boundary, leading to universal properties of the cosmological observables $n_s$ and $r$, identical to those of the universal attractor. This was subsequently generalized to the notion of $\alpha$-attractors \cite{Kallosh:2013yoa, Kallosh:2014rga}, building on the same idea but varying the \Kahler curvature and, with it, the tensor-to-scalar ratio, in the following way:
 \begin{align}
   R_K = -\frac{2}{3 \alpha}  \quad \Rightarrow \quad r = \frac{12 \alpha}{N^2} \,. \label{r}
 \end{align}
This remarkable relation between the \Kahler curvature and the tensor-to-scalar ratio was first found in \cite{Ferrara}. Finally, the non-minimal coupling and non-trivial \Kahler geometry can be related to each other and turn out to have a similar origin, that can be phrased as a pole of order two in the kinetic term of the inflaton \cite{Galante:2014ifa}. The residue of this pole is set by the parameter $\alpha$, which also has a beautiful interpretation \cite{Escher}.

In this Letter we present evidence for the universality of $\alpha$-attractors: while the previous models contain two chiral supermultiplets and employ a separation between the inflaton and the sGoldstino supersymmetry breaking directions \cite{Kawasaki:2000yn,Kallosh:2010xz}, we demonstrate that the same phenomenon can be achieved in a model containing just one superfield. The economical framework of realizing inflation in single-superfield models has been discussed in \cite{Goncharov:1985yu,Goncharov:1983mw,Achucarro:2012hg,Ketov:2014qha,Ketov:2014hya,Linde:2014ela,Linde:2014hfa,Kallosh:2015lwa}, but these do not include a variable \Kahler geometry and hence lack the parameter $\alpha$.

Our construction also highlights a novel approach to Minkowski and De Sitter model building.  Whereas the classic no-scale supergravity \cite{Cremmer:1983bf,Ellis:1983sf,Lahanas:1986uc} yields two flat Minkowski directions, one of these can be lifted to a stable direction by deforming the \Kahler curvature and allowing for a more general monomial dependence of the superpotential. Interestingly, a combination of these structures leads to a De Sitter plateau. This turns out to be stable only for such $\alpha$-deformed supergravities with a smaller \Kahler curvature than the one corresponding to a combination of no-scale constructions. Remarkably, generic deformations of these De Sitter plateaus lead to inflationary regimes with predictions \eqref{ns} and \eqref{r}.

Finally, analogous results emerge in the singular limit $\alpha\rightarrow\infty$ where the \Kahler geometry becomes flat. However, in this case the natural ingredients providing Minkowski or dS solutions and inflationary deformations will be exponentials, peculiar to this geometry.

\section{No-scale Supergravity and dS}

Our starting point will be the no-scale structure for a supergravity with a single chiral superfield. In this case the \Kahler potential reads
\be
K=-3 \ln\left(\Phi+\bar\Phi\right)\,,
\ee
describing a manifold SU(1,1)/U(1) and invariant under a shift of  Im($\Phi$), while the superpotential is independent of the superfield and hence constant. This model is characterized by a Minkowski vacuum in any point in field space. The negative definite contribution to the scalar potential, proportional to the square of the superpotential, is cancelled by the positive definite term, proportional to the square of the order parameter of supersymmetry breaking 
 \begin{align}
   D_\Phi W = \partial_\Phi W + K_\Phi W \,.
 \end{align}
Note that only the latter term of this contribution is non-vanishing due to the constancy of the superpotential. The resulting no-scale model necessarily has a flat direction along the imaginary part of $\Phi$, as this does not appear in either $K$ or $W$.

By a field redefinition, one can bring this simple no-scale model to a different form. In particular, in order to leave the \Kahler potential invariant, one can combine an inversion of the holomorphic field $\Phi$ with a specific \Kahler transformation, defined as
\be
\begin{aligned}\label{Ktranf}
K&\rightarrow K+ \lambda(\Phi) + \bar\lambda(\bar\Phi)\,, \quad
W \rightarrow \e^{- \lambda(\Phi)} W
 \end{aligned}
 \ee
with $\lambda = -3\ln\Phi$ in this case. Under these transformations, a constant superpotential becomes cubic instead. While this model has the same scalar potential and is therefore also of the no-scale type, it receives contributions from both terms in $D_\Phi W$.

Remarkably, one can combine the constant and cubic superpotentials to move away from no-scale models and generate a non-vanishing cosmological constant. In particular, the superpotential
  \begin{align}
    W = 1 - \Phi^3 \,,
\end{align}
 leads to a scalar potential with a flat direction along $\Phi=~\bar{\Phi}$, where the original Minkowski vacuum is shifted to $V =  \tfrac{3}{2}$ (while the combination with opposite sign leads to AdS). However, this De Sitter solution turns out to be unstable: the mass of the imaginary direction is given by $m_{\Im\Phi}^2 = - 2$.
 
\section{$\alpha$-scale Supergravity \& Stable dS}

In order to improve on the previous instability, we will consider a logarithmic \Kahler potential of the form
\be
K=-3\alpha\ln\left(\Phi+\bar\Phi\right)\,. \label{Kahler}
\ee
This still parametrizes a symmetric geometry $SU(1,1)/U(1)$, whose curvature is given by \eqref{r}.

\begin{figure}[htb]
\hspace{-3mm}
\begin{center}
\includegraphics[width=7cm,keepaspectratio]{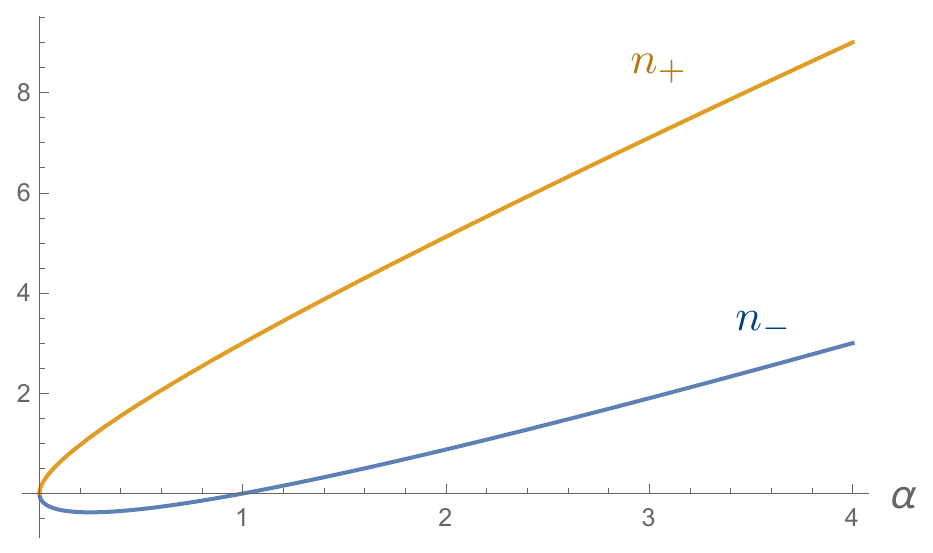}
\caption{The monomial powers $n_\pm$ for the $\alpha$-scale models as function of $\alpha$.}\label{nsol}
\end{center}
\vspace{-0.5cm}
\end{figure}

One can consistently truncate the equations of motion at $\Phi=\bar\Phi$ and, along this line, the relation between the geometric field and the canonical normalized field $\varphi$ is
\be
 \Phi= \bar \Phi = \e^{-\sqrt{\frac{2}{3\alpha}}\varphi}\,.
\ee
A  single monomial superpotential $W=\Phi^n$ will give a scalar potential equal to
\be
V=\frac{8^{-\alpha } \left[(2 n-3 \alpha )^2-9 \alpha \right] }{3 \alpha }\Phi ^{2 n-3 \alpha }\,,
\ee
along the real direction $\Phi=\bar\Phi$. Note that a constant potential corresponds to $2n=3 \alpha$ which, for any value of $\alpha$, leads always to AdS \cite{Ellis:2013nxa}. In contrast, a vanishing scalar potential corresponds to one of the following solutions
\be\label{npm}
n_\pm = \frac{3}{2} \left(\alpha \pm \sqrt{\alpha}\right)\,,
\ee
displayed in Fig.~\ref{nsol}. These are the counterparts of the constant and cubic superpotentials of the previous section, corresponding to $\alpha =1$. We will refer to the above model as {\it $\alpha$-scale supergravities} for the following reason. 

Similarly to the standard no-scale model, the real part of $\Phi$ has flat direction. On the other hand, the mass of the imaginary part gets a dependence on the field and, along $\Im\Phi=0$, reads
\be
m_{\Im\Phi}^2 =\frac{2^{2-3 \alpha } (\alpha -1) }{\alpha }\e^{\mp \sqrt{6}\varphi}\,,
\ee
where the sign of the power depends on the choice of one of the solutions \eqref{npm}. This result assures stability of the Minkowski vacuum for $\alpha\geq1$ \cite{GomezReino}.

Following the previous construction, one obtains a de Sitter plateau along the real direction by considering a pair of monomials,
\be\label{Wpol}
 W = \Phi^{n_-} - \Phi^{n_+}\,, \quad V= 3\cdot2^{2-3 \alpha } \,.
\ee
While this generically leads to terms with irrational powers, these are integers when $\alpha$ is a perfect square. Moreover, the more general choice with $9\alpha$ a perfect square yields still integer powers multiplied by an overall phase which can be gauged away by means of a \Kahler transformation.

Remarkably, unlike the standard case $\alpha=1$,  the mass of $\Im \Phi$ gets also some field dependent contributions and, along the real axis, reads
\be \label{dS-mass}
m_{\Im\Phi}^2 = -\frac{4 V}{3 \alpha} \left [ 1 - (\alpha -1) \sinh^2\left(\sqrt{\frac{3}{2}}\varphi\right)\right]\,.
\ee
Such a solution for the mass of the imaginary component allows to identify regions of stable de Sitter vacua. In particular, for $\alpha>1$, the field dependent terms dominate in the limit of large $|\varphi|$, leading to a positive mass for the imaginary component. Just a small region around the symmetric point ($\varphi=0$) leads to an instability, which disappears in the limit $\alpha\rightarrow\infty$.

\section{Single Superfield $\alpha$-attractors}

Once we know how to construct de Sitter in this context, we can add corrections to the superpotential \eqref{Wpol} in order to reproduce a consistent inflationary dynamics. Deviations from the positive plateau are given by higher powers $n$ of $\Phi$ ($n_- < n_+ < n$). In full generality, one can consider a deformation of the form
\be\label{Wgen}
W = \Phi^{n_-} - \Phi^{n_+} F(\Phi)\,,
\ee
with $F$ being a general function with an expansion $F(\Phi)=\sum_n c_n \Phi^n$. The corresponding scalar potential, in terms of the geometric field $\Phi$, reads
\be \label{potential}
V=\frac{2^{2-3 \alpha }  \left(\Phi F'(\Phi)+3 \sqrt{\alpha } F(\Phi)\right) \left( \Phi^{3 \sqrt{\alpha }+1} F'(\Phi)+3 \sqrt{\alpha }\right)}{3 \alpha }\,,
\ee
along the real axis, where primes denote derivatives with respect to $\Phi$. In the inflationary regime, close to $\Phi = 0$, only the first non-constant term is relevant: the scalar potential approximates
 \be\label{Vcan}
V= V_0  -  V_1 \e^{-\sqrt{\frac{2}{3\alpha}}\varphi} + \ldots \,,
\ee
at large values of the canonical field $\varphi$. The subleading terms will be irrelevant for order-one values of $\alpha$.

The inflationary scenario emerging from this construction is therefore the one typical of the $\alpha$-attractors: the \Kahler geometry, described by eq.~\eqref{Kahler}, determines unequivocally the observational predictions which, on the other hand, will be insensitive to specific changes in the superpotential.  Moreover, the predicted values for the spectral tilt and tensor-to-scalar ratio are \eqref{ns} and \eqref{r}, in the limit of large number of e-foldings $N$.

\begin{figure}[t!]
\hspace{-3mm}
\begin{center}
\includegraphics[width=7.5cm]{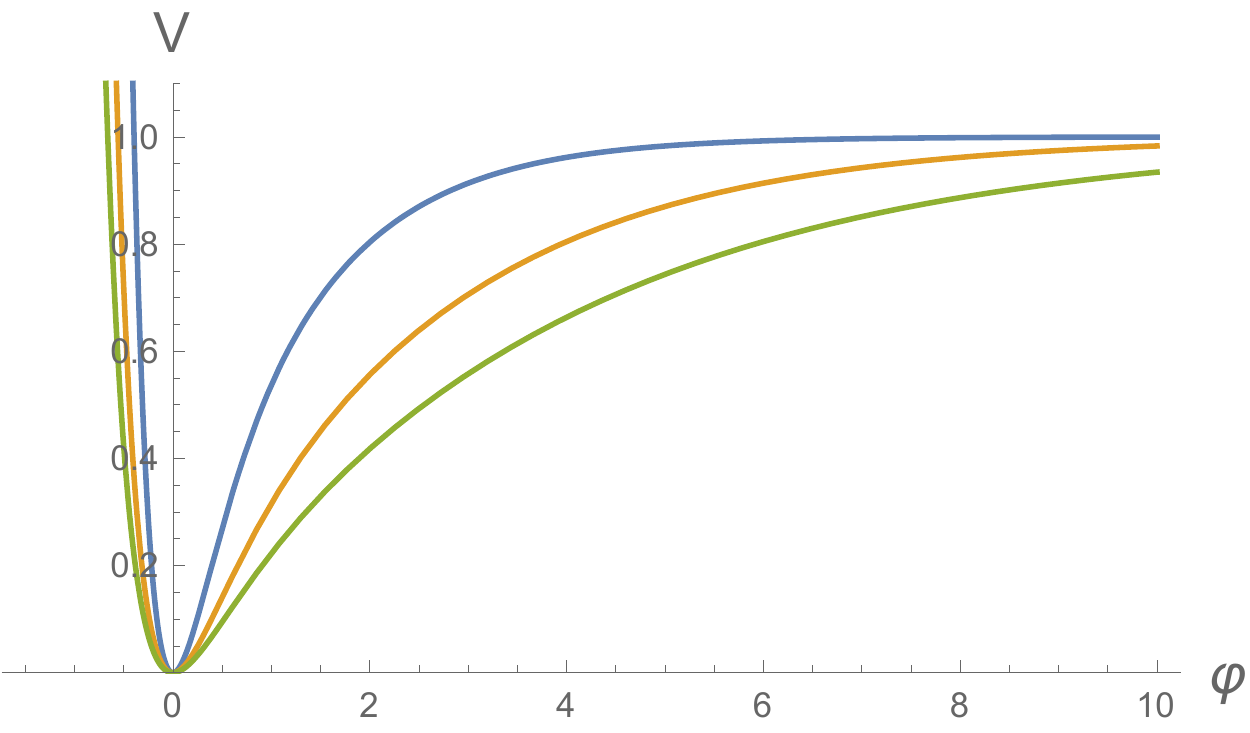}
\includegraphics[width=7.5cm]{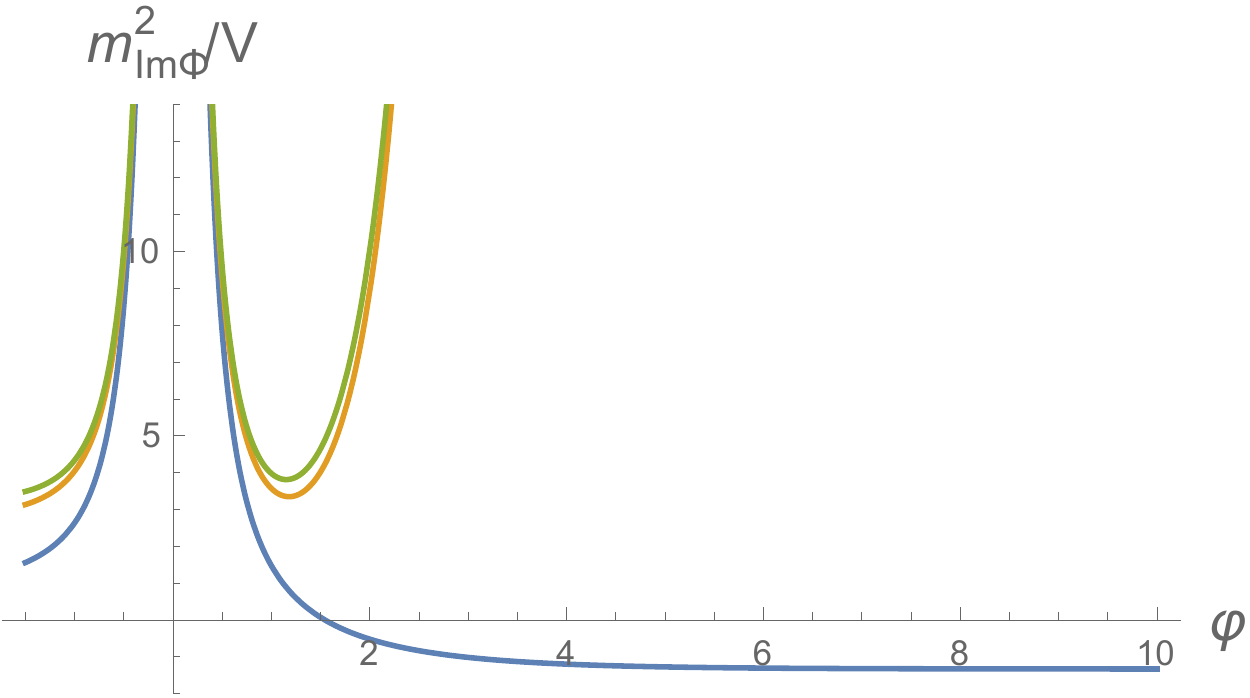}
\caption{Scalar potential and imaginary mass of the model defined by Eq.~\eqref{linear} in terms of $\varphi$ for $\alpha=\{1,4,9\}$. The blue line represents the instability occurring at $\alpha=1$.}\label{Vminimal}
\end{center}
\vspace{-0.75cm}
\end{figure}

\begin{figure}[t!]
\hspace{-3mm}
\begin{center}
\includegraphics[width=7.5cm]{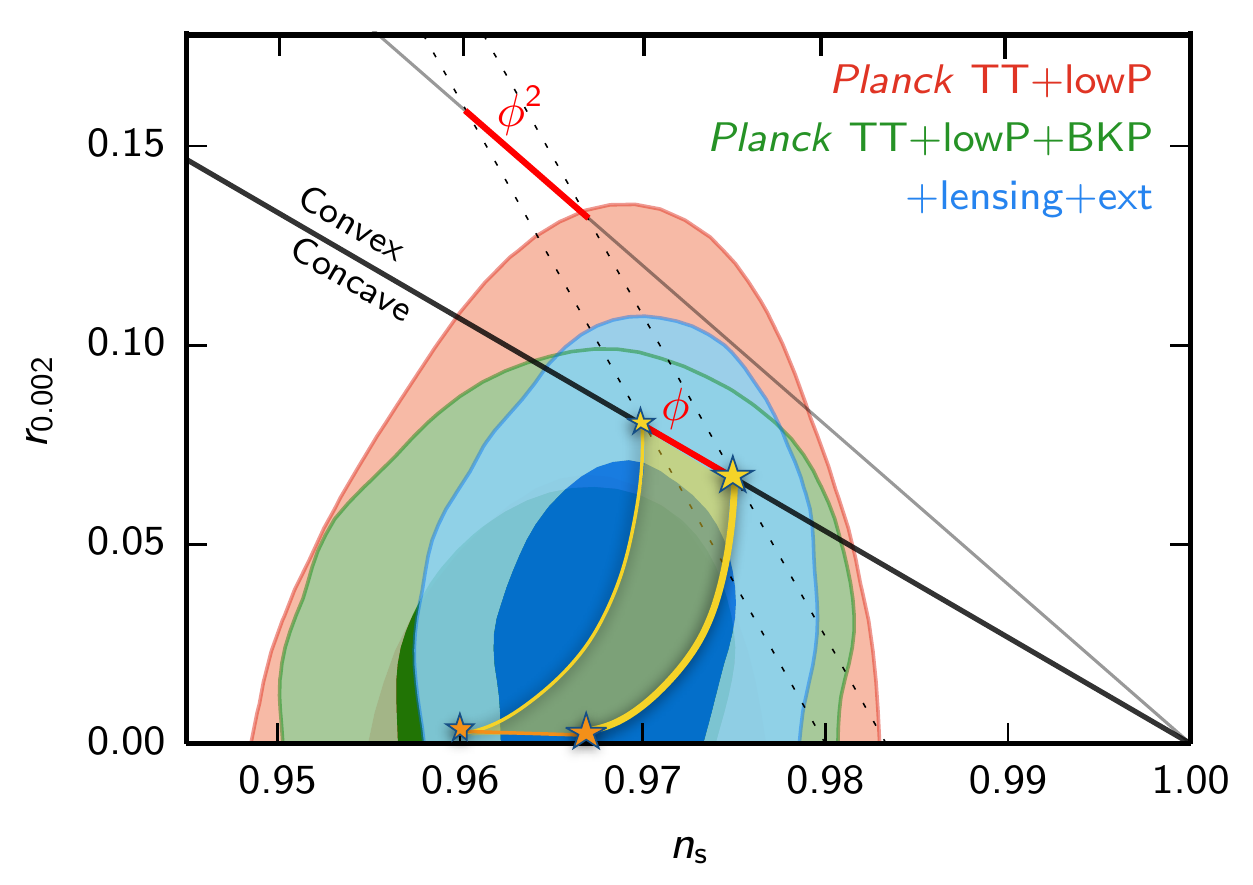}
\caption{The $(n_s ,r)$ predictions for $N=50$ and $N=60$ of the model Eq.~\eqref{linear} superimposed on the Planck constraints. The predictions interpolate between \eqref{ns} and \eqref{r} for small and order-one $\alpha$ and those of linear inflation for large $\alpha$.}\label{Predictions}
\end{center}
\vspace{-0.75cm}
\end{figure}

To demonstrate the stability and vacuum structure with an explicit example, we take
 \begin{align} \label{linear}
  F(x) = 1 + 3 \sqrt{\alpha} - 3 \sqrt{\alpha} x \,.
 \end{align}
These coefficients have been chosen to have a quadratic expansion around the Minkowski minimum at $\Phi = 1$. Both the scalar potential along the real axis as well as the mass of the imaginary direction is shown in Fig.~\ref{Vminimal} for different values of $\alpha$. This model is fully stable for $\alpha>1$ while the \Kahler curvature leads to an instability along the imaginary direction when  $\alpha\leq1$. Finally, its observational predictions superimposed on the confidence levels released by Planck2015 \cite{Planck:2015xua} are given in Fig.~\ref{Predictions}. These interpolate between the $\alpha$-attractor values \eqref{ns} and \eqref{r}, and those of a linear scalar potential.

The above approach leads to a supersymmetric Minkowski minimum. Uplifting this vacuum by means of supersymmetry breaking  to include a non-zero cosmological constant  is strongly constrained \cite{Kallosh:2014oja}: generically this cannot be done with a small deformation and, within one single superfield, leads to an undesirable large gravitino mass \cite{Linde:2014ela}. An additional nilpotent sector can elegantly solve the issue of the separation of the physical scales \cite{Kallosh:2014via,Dall'Agata:2014oka,Kallosh:2014hxa,Kallosh:2015lwa,Linde:2015uga, Scalisi:2015qga,CKL}. Nevertheless, as the two sectors prove to be independent from each other and play distinct roles \cite{Scalisi:2015qga}, it remains fundamental to construct a consistent inflationary dynamics in a single superfield context.

\section{Flat \Kahler limit}

In the singular limit $\alpha \rightarrow \infty$ the \Kahler geometry becomes flat. One could wonder whether there is a similar $\alpha$-scale model as well as de Sitter uplift in this limit. Indeed this is the case: upon a field redefinition $\Phi \rightarrow \exp( 2 \Phi / \sqrt{3 \alpha})$ and a \Kahler transformation with $\lambda = \tfrac{3}{2}\alpha\ln2 + \sqrt{3\alpha}\Phi$, the \Kahler potential \eqref{Kahler} yields
\be
K=-\tfrac{1}{2}\left(\Phi-\bar\Phi\right)^2,
\ee
in the singular limit $\alpha \rightarrow \infty$. Note that $K$ has become shift-symmetric in the inflaton field $\Re\Phi$ \cite{Kawasaki:2000yn}. This naturally provides a solution to the so-called $\eta$-problem \cite{Copeland:1994vg}, whereas, for finite values of $\alpha$, the latter is mitigated by the logarithmic form \eqref{Kahler} \cite{Roest:2013aoa}.

Under the same operations, the monomial superpotential turns into (modulo a constant, overall rescaling)
\be
W= \e^{\pm\sqrt{3} \Phi}\,.
\ee
One can check that this leads to a vanishing scalar potential along the line $\Phi = \bar \Phi$. Moreover, a linear combination of the two exponentials, such as  $W=\sinh(\sqrt{3} \Phi)$, leads to a constant and positive value of $V$ (while a cosh, instead, leads to AdS). The mass of the orthogonal imaginary component of $\Phi$ is equal to the $\alpha \rightarrow \infty$ limit of \eqref{dS-mass}.

The above construction can be perturbed to have deviations from de Sitter and produce a consistent inflationary dynamics. A first guess could be to include the same deformation in the polynomial \eqref{Wgen} and take the $\alpha \rightarrow \infty$ limit. However, in this case the field dependence of this function is washed out: for finite values of the constants $c_n$, the resulting superpotential reads 
 \begin{align}
  W = \e^{\sqrt{3} \Phi} - \e^{- \sqrt{3} \Phi} F(1) \,,
 \end{align}
 leading to a constant scalar potential.
 
A more natural possibility, given the exponential ingredients of the above superpotential, would be to take
 \be\label{Wdev}
   W= \e^{\sqrt{3} \Phi} - \e^{-\sqrt{3} \Phi} F \left(\e^{- 2 \Phi / \sqrt{3 \alpha'}}\right) \,,
 \ee
where we have parametrized the additional dependence in terms of a new parameter $\alpha'$. Remarkably, when truncating to the real axis, the scalar potential arising from this supergravity model with a flat \Kahler geometry is identical to \eqref{potential} of the supergravity model with a curved \Kahler geometry, provided one identifies $\alpha = \alpha'$. The specific choice \eqref{linear} for $F$ in this case leads to the identical predictions of Fig.~3; however, interestingly, this model proves to be stable for any positive value of $\alpha$.
 
It therefore turns out to be possible to represent the same single-field inflationary potential by means of curved or flat \Kahler geometry. Only the former has the attractive interpretation of the robustness of $\alpha$-attractors arising from a non-trivial \Kahler geometry; the same dynamics arises in the flat case by the peculiar non-polynomial form of $W$. 

The first example of such a model \cite{Goncharov:1985yu,Goncharov:1983mw,Linde:2014hfa} fits perfectly into the recipe given above: it has a superpotential 
 \begin{align}
  W = \sinh(\sqrt{3} \Phi)\tanh(\sqrt{3} \Phi)\,, 
  \end{align}
corresponding to the choice  $F(x) = (3- x)/(1+x)$ for the case of $\alpha' = 1/9$. The same inflationary potential can also be embedded in a logarithmic \Kahler structure \cite{Kallosh:2015lwa}.

\section{Discussion}\label{disc}

In this Letter we have outlined a strikingly simple route to construct single superfield models with stable de Sitter solutions. Generic deformations of these models yield an inflationary trajectory fully consistent with Planck. The key quantity in this set of models, similar to the original $\alpha$-attractors, is the curvature of the \Kahler manifold \eqref{r}. This quantity determines both the (in)stability of such constructions as well as the inflationary predictions of the deformed models. 

Remarkably, this provides a realization of $\alpha$-attractors employing a single superfield, in constrast to the two-field model of \cite{Kallosh:2013yoa,Kallosh:2014rga}.
This suggests that the phenomenon of \Kahler curvature leading to the inflationary predictions \eqref{ns} and \eqref{r} is universal, and applies to a much larger set of \Kahler geometries than $SU(1,n) / U(n)$ with $n=1,2$. 

Given the prominence of no-scale models in the literature, it would be interesting to study other possible applications of the $\alpha$-generalization proposed in this Letter. An example could be the no-scale inflationary constructions of \cite{Ellis:2013xoa,Ellis:2013nxa,Ellis:2014gxa,Ellis:2014opa,Lahanas:2015jwa}. Moreover, while we have focused on single superfield models, it is straightforward to generalize this construction to multi-fields:
 \begin{align}
  K = \sum_i - 3 \alpha_i \log(\Phi_i + \bar \Phi_i) \,,  \quad
   W = \prod_i \Phi_i^{n_i} \,, \label{multi}
 \end{align}
where we have suppressed other fields with a different dependence. The condition for Minkowski is
 \begin{align}
 \sum_i \frac{(2n_i - 3 \alpha_i)^2}{3\alpha_i} = 3 \,.
  \end{align}
Remarkably, also in the multi-field case, the interference of superpotential terms with flat Minkowski vacua leads to a de Sitter phase, proving the generality of such a feature. It would be very interesting to investigate the stability and inflationary aspects of such constructions.
  
Finally, our construction invites investigations of string theory scenarios leading to \eqref{multi}. Many moduli contribute with a factor $\alpha_i = 1/3$ to the \Kahler potential, while flux compactifications yield polynomial contributions to the superpotential. It would be of the utmost interest to realize this in a concrete setting.



\section*{Acknowledgments}

We thank Dries Coone, Renata Kallosh, Andrei Linde and Pablo Ortiz for very stimulating discussions as well as valuable comments on a draft of this paper.

\bibliography{refsAS}
\bibliographystyle{utphys}

\end{document}